\documentclass[twocolumn,preprintnumbers,amsmath,amssymb]{revtex4}
\usepackage{graphicx}
\usepackage[ansinew]{inputenc}
\usepackage{color}

\def\be{\begin{equation}}
\def\ee{\end{equation}}
\def\bea{\begin{eqnarray}}
\def\eea{\end{eqnarray}}
\newcommand{\ket}[1]{\mbox{$|#1\rangle$}}
\newcommand{\bra}[1]{\mbox{$\langle#1|$}}

\begin{document}

\bibliographystyle{naturemag}

\title{The origin of quantum nonlocality }

\author{Fang-Yu Hong}

\affiliation{Department of Physics, Center for Optoelectronics Materials and Devices, Zhejiang Sci-Tech University, Xiasha College Park, Hangzhou, Zhejiang 310018, CHINA
}

\date{\today}

\begin{abstract}
Quantum entanglement is the quintessential characteristic of quantum mechanics and the basis for quantum information processing. When one of two maximally entangled particles is measured, without measurement the state of another one is determined simultaneously no matter how far the two particles is from each other.  How can these phenomena take place since no object can move faster than light speed in a vacuum?  The key problem is due to the ignorance of the interaction between a particle and a quantum vacuum. Just like the case where a gun suffers recoil from  its firing of a bullet, when a particle is created from the quantum vacuum, the vacuum will be somewhat ``broken" correspondingly, which can be described by a shadow state in the vacuum.  Through their shadows in the vacuum two quantum entangled particles can have a distance-independent instantaneous interaction with each other. Quantum teleportation, quantum swap, and wave function collapse are explained in a similar way. Quantum object can be interpreted as a composite made up of a particle and the shadowed quantum vacuum which is responsible for the wave characteristic of the particle wave duality.   The quantum vacuum is not only the origin of all possible kinds of particles, but also the origin and the core of Eastern mystics.
\end{abstract}


\maketitle

 In spite of its unprecedented success in description of nature,  quantum theory  remains mysterious.  Even great physicist like Feynman, who presented a new formulation of quantum mechanics, at one time said  that nobody understands quantum mechanics. This statement becomes clear when one realized that quantum theory is based on a number of very counterintuitive concepts and notions and many questions such as those about the nature of the external world and the meaning of an objective physical reality remain open.  Classical physics impresses us many deep-seated concepts such as   'realism'
according to which an external reality exists independent of
observation and 'locality' which means that local events cannot be
affected by actions in space-like separated regions \cite{jfcs}.  Based on
these  deep-rooted reasonable assumptions, in 1935, Einstein, Podolsky and Rosen (EPR) originated
the famous ``EPR paradox". The EPR paper
spurred intensive  investigations into the nonlocality of quantum physics.  Furthermore,
the EPR paradox brought into sharp focus the
concept of entanglement, now considered to be the cornerstone
of quantum information process.\cite{aepb}.  Since 1964 the EPR arguments about the
physical reality of quantum systems is shifted from the realm of
philosophy to the domain of experimental physics  when
Bell and others constructed mathematical inequalities - one of the
profound scientific discoveries of the 20th century \cite{jsb,jfch},
which must be satisfied by any theory based on the joint assumption
of realism and locality and may be violated by quantum mechanics.
Many experiments \cite{jfcs,sjfc,esft,aagp,zyom,prtr,pgkm,wtbj,mard, dsab,dmpm}
have since been done that are consistent with quantum mechanics and
inconsistent with local realism. All of the present experimental evidence leads to the conclusion that quantum correlations take place instantaneously  and nature is nonlocal\cite{gisin}. Then there arises a big question: why quantum correlations happen instantaneously as if the space collapses for two entangled particles at a far distance?

Here I show that two entangled particles can interact distance-independently through in the quantum vacuum (QV) the shadows which accompany the generation of the particle and change with the particle states. Quantum teleportation \cite{dbjp}, quantum swap \cite{jpdb}, and wave function collapse can be explained in a similar way. A quantum object is interpreted as a composite consisting of two inseparable components - a particle and the shadowed QV which is responsible for wave characteristic of the so-called particle wave duality. The QV is not only the origin of all kinds of particles, but also the same origin of the counterintuitive concepts in quantum physics and the mystery in  Eastern mysticism \cite{capra}.

\begin{figure}[t]
\includegraphics[width=8cm]{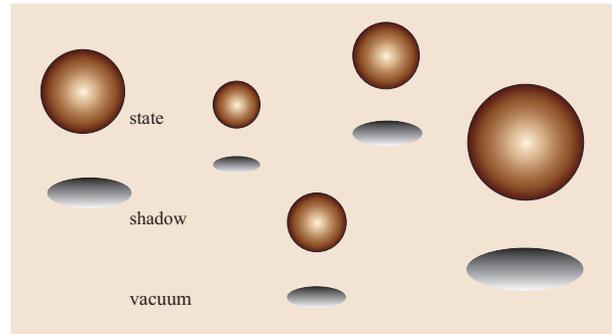}
\caption{\label{fig1} Picture of quantum objects. The QV is the origin of all kinds of particles. When a particle is created from a QV, the QV is in some sense ``broken", which can be described by a shadow state in the QV.  The shadow state and the particle state have an one-to-one mapping; the shadow state changes correspondingly with the particle state. The QV is the physical reality containing all possible kinds of quantum fields which are all in vacuum states. Any quantum object is the composite of two inseparate components, a particle and the shadowed QV. It is the QV that makes a quantum object have a wave nature.}
\end{figure}

First we discuss a boson-type quantum field. A boson of momentum ${\bf p}_j$ can be created from a QV
\be\label{eq1}
\ket{1}_{{\bf p}_j}=a({{\bf p}_j})^\dagger\ket{0}  \ee
where $a({\bf p})^\dagger$ denotes the creation operator (Fig. \ref{fig1}). The QV is the physical reality comprising of all possible quantum fields which are all in the vacuum states.  This  equation can be rewritten as follows
 \bea\label{eq17}
\ket{1}_{{\bf p}_j}&\rightarrow&\ket{1}_{{\bf p}_j}^{qv}=a({\bf p}_j)^\dagger\ket{\cdots0_{{\bf p}_i}\cdots 0_{{\bf p}_j}\cdots0_{{\bf p}_k}\cdots}\, \notag \\
&=&\ket{1}_{{\bf p}_j}\ket{\cdots0_{{\bf p}_i}\cdots \bar{0}_{{\bf p}_j}\cdots0_{{\bf p}_k}\cdots}\notag\\
&\equiv&\ket{1}_{{\bf p}_j}\bigcirc\ket{ \bar{1}_{{\bf p}_j}}
\eea
where the superscript ${qv}$ denotes the state of the system comprising a particle, the QV and the corresponding  shadow in the QV, $\ket{\bar{0}_{{\bf p}_j}}$ describes the absence of the vacuum state $\ket{0_{{\bf p}_j}}$, $\bigcirc\equiv\ket{\cdots0_{{\bf p}_i}\cdots 0_{{\bf p}_j}\cdots0_{{\bf p}_k}\cdots}$ denotes the QV state  and $\ket{ \bar{1}_{{\bf p}_j}}$ named  the shadow state of $\ket{ 1_{{\bf p}_j}}$ in the quantum vacuum
describes the effect of the creation of a particle with momentum ${\bf p}_j$ on the vacuum. The physical meaning of  Eq.\eqref{eq1} is obvious: when a particle of momentum ${\bf p}_j$  is created from the QV, the QV changes correspondingly by leaving the quantum vacuum mode $\ket{0_{{\bf p}_j}}$ absent - the QV is somewhat ``broken", in the similar way as a hole will appear in a valence band when an electron is excited from the valence band to a conduction band. Thus the generation of any particle from the QV will leave in the QV a ``hole" named a shadow and  any quantum state will have a corresponding shadow state in the QV.  How can we expect that there isn't any effect on the QV when a particle is created from the body of the QV and evolves  in the body of the QV? It is very regretable that so far we have all through neglected this crucial effect. Here this effect is considered and described by the shadow of a particle and the shadow state of a particle state.

From equation \eqref{eq17} we see that the generation of a particle of momentum ${\bf p}$  from the QV will leave a corresponding shadow state in the QV. It is natural that the generation of $n$ particles with momentum ${\bf p}$ will leave a corresponding shadow state in the QV too. To describe this situation, I introduce a shadow operator $\bar{a}^\dagger({\bf p})$ which acts on the shadow state, then Eq. \eqref{eq17} can be rewritten as
\be\label{eq18}
\ket{1}_{{\bf p}_j}^{qv}=a({\bf p}_j)^\dagger \bar{a}({\bf p}_j)^\dagger\bigcirc
\equiv\ket{1}_{{\bf p}_j}\bigcirc\ket{ \bar{1}_{{\bf p}_j}}
\ee
The action of  annihilation operators $ a({\bf p})$ and $\bar{a}({\bf p})$  on the QV is written as
\be\label{eq19}
a({\bf p}_j)\bar{a}({\bf p}_j)\bigcirc=0\bigcirc,
\ee
which has stressed the fact that the QV can't be eliminated. Since any quantum state always accompanies  its shadow state, the annihilation operator $ a({\bf p})$ and $\bar{a}({\bf p})$ or the creation operation $a({\bf p}_j)^\dagger $ and $\bar{a}({\bf p}_j)^\dagger$ should always appear together. Any operator and its shadow operator are commutable with each other; the shadow operators obey the same commute relations as the corresponding  operators.

We can express the combination of the operators $ a({\bf p})\bar{a}({\bf p})$ by $ b({\bf p})$,
\be\label{eq20}
b({\bf p})^\dagger\bigcirc=\ket{1}_{{\bf p}_j}\bigcirc\ket{ \bar{1}_{{\bf p}_j}}
\ee
and
\be\label{eq35}
b({\bf p})^\dagger\ket{n}_{{\bf p}_j}\bigcirc\ket{ \bar{n}_{{\bf p}_j}}=\sqrt{n+1}\ket{n+1}_{{\bf p}_j}\bigcirc\sqrt{n+1}\ket{ \overline{n+1}_{{\bf p}_j}}.
\ee
The commutation relations are
\bea\label{eq25}
&&[b({\bf p}), b^\dagger({\bf p'})] =\delta^{(3)}({\bf p}-{\bf p'});\notag\\
&&[b({\bf p}), b({\bf p'})] =0;[b^\dagger({\bf p}), b^\dagger({\bf p'})]=0.
\eea
For  Fermions, the commutation relations have to be replaced by anticommutation relations.

In the Heisenberg picture, any operator $O$ changes with time according to the equation of motion
\be \label{eq22}
i\hbar\frac{\partial}{\partial t}O=[O, H(b,b^\dagger)],
\ee
through which any interactions between particles or the external field can be introduced \cite{mpds}.
The field operator
\be\label{eq21}
\phi(x)=\int \frac{d^3p}{(2\pi)^{3/2}\sqrt{2E_p}}(b_pe^{-ip\cdot x}+b_p^\dagger e^{ip\cdot x})
\ee
acting on the QV, creates a particle of mass $m$ at position ${\bf x}$ and a corresponding shadow in the QV \cite{mpds}
\be\label{eq28}
\phi(x)\bigcirc=\int\frac{d^3p}{(2\pi)^{3/2}\sqrt{2E_p}} e^{ip\cdot x}\ket{{\bf p}}\bigcirc\ket{\bar{{\bf p}}}.
\ee
Here $p\cdot x= E_pt-{\bf p}\cdot{\bf x}$ with $E_p=\sqrt{p^2c^2+m^2c^4}$. For the case of nonrelativistic  particles,  the field operator is expanded as follows
\be\label{eq29}
\phi({\bf x},t)=\int \frac{d^3p}{(2\pi)^{3/2}}b_pe^{i{\bf p}\cdot{\bf x}-i\frac{p^2}{2m}t}.
\ee
$\phi({\bf x},t)^\dagger$ acting on the QV
\be\label{eq23}
\ket{{\bf x},t}\bigcirc\ket{\bar{{\bf x}},\bar{t}}=\phi({\bf x},t)^\dagger\bigcirc
\ee
  creates a particle of mass $m$ at position ${\bf x}$ and a corresponding shadow in the QV. Wave function is defined as
  \bea\label{eq24}
\psi({\bf x},t)\Big|\bar{\psi}({\bf x},t)&=&\bra{\cdots,\bar{n}_2,\bar{n}_1}\bigcirc\bra{\cdots,n_2,n_1}{\bf x},t\rangle\bigcirc\ket{\bar{{\bf x}},\bar{t}}\notag\\
&&(n_1+n_2+\cdots =1),
\eea
where the notation $\Big|$ stresses the fact that $\bar{\psi}({\bf x},t)$ is the shadow state of $\psi({\bf x},t)$ and
\be\label{eq26}
\ket{n_1,n_2,\cdots}\bigcirc\ket{\bar{n}_1,\bar{n}_2,\cdots}=(b^\dagger({\bf p}_2))^{n_1}(b^\dagger({\bf p}_1))^{n_2}\cdots\bigcirc.
\ee
If a particle of mass $m$ is subject to an external field $V({\bf x})$, the corresponding Hamiltonian  has the form
\be\label{eq27}
H=\int d^3x\left[-\frac{1}{2m}\phi^\dagger\nabla^2\phi+\phi^\dagger V({\bf x})\phi\right].
\ee
According to the Heisenberg equation of motion Eq.\eqref{eq22}, we have \cite{luri}
\be\label{eq30}
i\hbar\frac{\partial}{\partial t}\phi({\bf x},t)=\left(-\frac{1}{2m}\nabla^2+V({\bf x})\right)\phi({\bf x},t).
\ee
From Eqs. (\ref{eq23},\ref{eq24},\ref{eq30}), we obtain
\be\label{eq31}
i\hbar\frac{\partial}{\partial t}\psi({\bf x},t)\Big|\bar{\psi}({\bf x},t)=\left(-\frac{1}{2m}\nabla^2+V({\bf x})\right)\psi({\bf x},t)\Big|\bar{\psi}({\bf x},t).
\ee
This equation just tell us that the wave function $\psi({\bf x},t)$ obeys the Schr\"{o}dinger equation
\be\label{eq32}
i\hbar\frac{\partial}{\partial t}\psi({\bf x},t)=\left(-\frac{1}{2m}\nabla^2+V({\bf x})\right)\psi({\bf x},t)
\ee
while the shadow wave function $ \bar{\psi}({\bf x},t)$  obeys the same equation of motion as $\psi({\bf x},t)$.

 Thus any quantum state $\ket{\psi}$ should have a corresponding shadow state $\ket{\bar{\psi}}$ in the QV, leading to the quantum object should be described by $\ket{\psi}^{qv}=\ket{\psi}\bigcirc\ket{\bar{\psi}}$. For any superposition state $\ket{\psi}=\sum_i c_i\ket{\phi_i}$, the corresponding particle-vacuum state should be written by
\bea\label{eq2}
 \ket{\psi}^{qv}&=&\sum_i c_i\ket{\phi_i}\bigcirc\sum_i c_i\ket{\bar{\phi}_i}\notag\\
 &\equiv&\sum_ic_i(\ket{\phi_i}\bigcirc\ket{\bar{\phi_i}})\equiv\sum_ic_i\ket{\phi_i}\bigcirc\ket{\bar{\phi_i}}
\eea
where $c_i$ is a coefficient. This equation can be easily obtained from the fact that the $\ket{\bar{\psi}}(\ket{\bar{\phi_i}})$ is always the shadow of the state $\ket{\psi}(\ket{\psi_i})$ . To simplify  notations, I apply the last expression in Eq.\eqref{eq2}.  From this equation, we can see that the particle-vacuum system can't be seen as an ordinary  system comprising  two particles. Considering there exist many different kinds of fields, the QV should be described by
\bea\label{eq5}
\bigcirc &=&\cdots\ket{\cdots0_{{\bf p}_i}^{s_m}\cdots0_{{\bf p}_j}^{s_m}\cdots0_{{\bf p}_k}^{s_m}\cdots}\notag \\
&\cdots&\ket{\cdots0_{{\bf p}_i}^{s_n}\cdots0_{{\bf p}_j}^{s_n}\cdots0_{{\bf p}_k}^{s_n}\cdots}\cdots
\eea
where $s_i$ represent the quantum numbers other than momentum ${\bf p}$.

\begin{figure}[t]
\includegraphics[width=8cm]{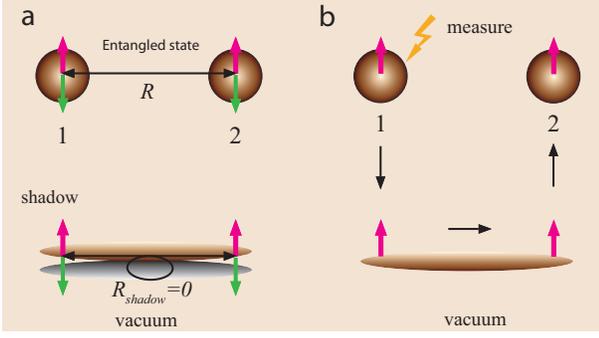}
\caption{\label{fig2} The QV-based interaction between two entangled particles. (a) Two electron spins are in the maximally entangled state $\ket{\phi^+}$, which has a corresponding shadow state $\ket{\bar{\phi}^+}$ in the QV. The state of the system can be described by $\ket{\phi^+}\bigcirc\ket{\bar{\phi}^+}$. The two entangled electron spins interacts with each other through the QV and the shadow state $\ket{\bar{\phi}^+}$. Note that this interaction is independent of the distance between the electrons and is instantaneous, as if in the QV  the distance  $R_{shadow}=0$. (b)  The  state of spin 1 is read out, e.g., to be $\ket{\uparrow}_1$. this readout will simultaneously cause  the shadow state  changes from $\ket{\bar{\phi}^+}$ into $\ket{\bar{\uparrow}}_1\ket{\bar{\uparrow}}_2$, leading to simultaneously  the spin state  $\ket{\uparrow}_2$. The readout of the spin state 1, the change of the shadow state, and the determination of the second spin state are simultaneous through the QV where the coordinate space seems to have collapsed.}
\end{figure}

When two particles are entangled, e.g., two electron's spins are maximally entangled (Fig. \ref{fig2})
\be\label{eq3}
\ket{\phi^+}=\sqrt{2}\left(\ket{\uparrow}_1\ket{\uparrow}_2+\ket{\downarrow}_1\ket{\downarrow}_2\right).
\ee
Considering their shadows, the state of the system considered should be written as
\bea\label{eq4}
\ket{\phi^+}^{qv}&=&\ket{\phi^+}\bigcirc\ket{\bar{\phi}^+}=
\frac{1}{\sqrt{2}}\left(\ket{\uparrow}_1\ket{\uparrow}_2\bigcirc\ket{\bar{\uparrow}}_1\ket{\bar{\uparrow}}_2\right.\notag\\
&+&\left.\ket{\downarrow}_1\ket{\downarrow}_2\bigcirc\ket{\bar{\downarrow}}_1\ket{\bar{\downarrow}}_2\right).
\eea
When a readout of spin 1 is performed with a result $\ket{\uparrow}_1$, the shadow state in the QV becomes $\bigcirc\ket{\bar{\uparrow}}_1\ket{\bar{\uparrow}}_2$, which shows that the state of spin 2 is $\ket{\uparrow}_2$. The determination of the state of  particle 1, the shadow state, and that of particle 2 is taken place simultaneously. The interaction between particle 1,2 based on the QV through the shadow state is well described in the Fock space (Eq. \eqref{eq4}) and is independent of the distance between the two particles. If the interaction is described  in the point of view of the coordinate space, the interaction takes place with an infinity speed, or the distance between the particles  is taken as zero, as if the space collapses in the QV. This explain of the nonlocal characteristic of the entanglement is very natural, because the QV exists  everywhere, and the two particles exist in the QV, Once more I stress that the QV-based influence between the two particles takes place without any time-ordering,   which is in consistent with the results in literatures \cite{dsab,gisin, ashz, hzjb}.

As a contrast, we consider the case where the two particles is not entangled and is described  by
\bea\label{eq6}
\ket{\psi}^{qv}&=&\frac{1}{2}
\left(\ket{\uparrow}_1+\ket{\downarrow}_1\right)\left(\ket{\uparrow}_2+\ket{\downarrow}_2\right)\notag\\
&\bigcirc&\frac{1}{2}\left(\ket{\bar{\uparrow}}_1+\ket{\bar{\downarrow}}_1\right)\left(\ket{\bar{\uparrow}}_2+
\ket{\bar{\downarrow}}_2\right).
\eea
When the spin state of  particle 1 is read out with a result $\ket{\uparrow}_1$, the shadows in the QV turn into
$\bigcirc\ket{\bar{\uparrow}}_1\frac{1}{\sqrt{2}}\left(\ket{\bar{\uparrow}}_2+\ket{\bar{\downarrow}}_2\right)$
which shows that particle 2 is in the state $\frac{1}{\sqrt{2}}\left(\ket{\uparrow}_2+\ket{\downarrow}_2\right)$. Thus the detection of particle 1 has no influence on particle 2, though the quantum information of the detection result of particle 1 transfers instantaneously  to every place of the coordinate space.

 Next we consider the teleportation  of an unknown state. Alice have two electrons 1 and 2, and Bob has electron 3. The spin of electron 1  is in an unknown state $\alpha\ket{\uparrow}_1+\beta\ket{\downarrow}_1$ with $|\alpha|^2+|\beta|^2=1$, electrons 2,3 are in the maximally entangled state
\be\label{eq7}
\ket{\psi^-}=\frac{1}{\sqrt{2}}
\left(\ket{\uparrow}_2\ket{\downarrow}_3-\ket{\downarrow}_2\ket{\uparrow}_3\right).
\ee
Taking into the shadows in the QV, the state of the considered system has the form
\bea\label{eq8}
&&\ket{\psi}^{qv}_{123}=(\alpha\ket{\uparrow}_1+\beta\ket{\downarrow}_1)\frac{1}{\sqrt{2}}
\left(\ket{\uparrow}_2\ket{\uparrow}_3-\ket{\downarrow}_2\ket{\downarrow}_3\right)\notag\\
&&\bigcirc (\alpha\ket{\bar{\uparrow}}+\beta\ket{\bar{\downarrow}})\frac{1}{\sqrt{2}}
\left(\ket{\bar{\uparrow}}_2\ket{\bar{\uparrow}}_3-\ket{\bar{\downarrow}}_2\ket{\bar{\downarrow}}_3\right)
\eea
This state can be rewritten as
\bea\label{eq9}
&&\ket{\psi}^{qv}_{123}=\frac{1}{2}[\ket{\psi^-}_{12}(-\alpha\ket{\uparrow}_3-\beta\ket{\downarrow}_3)\notag\\
&&+\ket{\psi^+}_{12}(-\alpha\ket{\uparrow}_3+\beta\ket{\downarrow}_3)
+\ket{\phi^-}_{12}(\alpha\ket{\uparrow}_3+\beta\ket{\downarrow}_3)\notag\\
&&+\ket{\phi^+}_{12}(\alpha\ket{\uparrow}_3-\beta\ket{\downarrow}_3)]\bigcirc
\frac{1}{2}[\ket{\bar{\psi}^-}_{12}(-\alpha\ket{\bar{\uparrow}}_3-\beta\ket{\bar{\downarrow}}_3)\notag\\
&&+\ket{\bar{\psi}^+}_{12}(-\alpha\ket{\bar{\uparrow}}_3+\beta\ket{\bar{\downarrow}}_3)
+\ket{\bar{\phi}^-}_{12}(\alpha\ket{\bar{\uparrow}}_3+\beta\ket{\bar{\downarrow}}_3)\notag\\
&&+\ket{\bar{\phi}^+}_{12}(\alpha\ket{\bar{\uparrow}}_3-\beta\ket{\bar{\downarrow}}_3)]
\eea
with the Bell states
\be\label{eq10}
\ket{\psi^\pm}_{12}=\frac{1}{\sqrt{2}}(\ket{\uparrow}_1\ket{\downarrow}_2\pm\ket{\downarrow}_1\ket{\uparrow}_2)
 \ee
 and
 \be\label{eq11}
\ket{\phi^\pm}_{12}=\frac{1}{\sqrt{2}}(\ket{\uparrow}_1\ket{\uparrow}_2\pm\ket{\downarrow}_1\ket{\downarrow}_2).
 \ee
Alice makes a Bell-state measurement on the spins of two particles 1 and 2, and find particles 1 and 2 is in one of the four Bell states, e.g. $\ket{\psi^-}_{12}$, leading to the corresponding  shadow state $\bigcirc
\ket{\bar{\psi}^-}_{12}\frac{1}{\sqrt{2}}(-\alpha\ket{\bar{\uparrow}}_3-\beta\ket{\bar{\downarrow}}_3)$,  which determines particle 3 is in the state $\frac{1}{\sqrt{2}}(-\alpha\ket{\uparrow}_3-\beta\ket{\downarrow}_3)$. Alice tells Bob his measurement result through a classic channel, then Bob carries out a corresponding  unitary transform on particle 3 leaving it in state $\frac{1}{\sqrt{2}}(\alpha\ket{\uparrow}_3+\beta\ket{\downarrow}_3)$. Thus the quantum teleportation of an unknown state is performed with the aid of the QV and the shadow states.

Quantum entanglement can be connected through quantum swap. The spins of four electrons  are initialized in the entangled states $\ket{\psi^-}_{12}$ and $\ket{\psi^-}_{34}$. Considering the QV and the shadows states the system can be described by
\be\label{eq12}
\ket{\psi}^{qv}_{1234}=\ket{\psi^-}_{12}\ket{\psi^-}_{34}\bigcirc\ket{\bar{\psi}^-}_{12}\ket{\bar{\psi}^-}_{34}
\ee
This state can be rewritten as follows
\bea\label{eq13}
&&\ket{\psi}^{qv}_{1234}=\frac{1}{2}[\ket{\psi^+}_{14}\ket{\psi^+}_{23}-\ket{\psi^-}_{14}\ket{\psi^-}_{23}\notag\\
&&-\ket{\phi^+}_{14}\ket{\phi^+}_{23}+\ket{\phi^-}_{14}\ket{\phi^-}_{23}]\bigcirc\frac{1}{2}[\ket{\bar{\psi}^+}_{14}\ket{\bar{\psi}^+}_{23}\notag\\
&&-\ket{\bar{\psi}^-}_{14}\ket{\bar{\psi}^-}_{23}-\ket{\bar{\phi}^+}_{14}\ket{\bar{\phi}^+}_{23}+\ket{\bar{\phi}^-}_{14}\ket{\bar{\phi}^-}_{23}]
\eea
 Alice has two electrons 2 and 3, and Bob has the others. Alice carries out a Bell-state measurement on the spins 2 and 3, and finds them is in one of four Bell states, e.g. $\ket{\psi^+}_{23}$, which guarantees that the corresponding  shadow state is $\ket{\bar{\psi}^+}_{14}\ket{\bar{\psi}^+}_{23}$. This shadow state ensures that  electrons 1 and 4 are in the state $\ket{\psi^+}_{14}$. The determination of the state of electrons 2 and 3, the shadow state of electrons (2, 3)  and  (1, 4), and the state of electrons 1 and 4 are simultaneous, though Bob doesn't know the state of the electrons at his side until he receives the classic state information from Alice.   Here we can see that quantum swap is also performed under  the help of the  QV and the shadow states in the QV.

Now we discuss the collapse of wave functions in the whole coordinate space. A particle is prepared in a state $\psi({\bf r},t)$. The space can be divided into infinity zones labeled by $i (i=1,2,\cdots)$ with infinitesimal volume $dV_i$. Then this state can be expanded as follows
\be\label{eq14}
\psi({\bf r},t)=\sum_{i}c_i\varphi^i
\ee
where coefficient $c_i$ fulfill $\sum_i|c_i|^2=1$  and wave function $\varphi^j$  comprises infinity parts
\be\label{eq15}
\varphi^j=\cdots\circ\phi_i^j\circ\cdots\phi_j^j\circ\cdots\phi_k^j\circ\cdots.
\ee
  Here wave function $\phi_i^j$ is  defined only in  zone $i$ of volume $dV_i$ and has $|\phi_j^j|^2dV_i=1$ and $\phi_i^j=0 \,\text {for}\, i \neq j $. Combined with the shadow state in the QV, the state of the considered system  can be written as
 \be\label{eq116}
\psi({\bf r},t)=\sum_{i}\varphi^i\Big\vert\sum_{i}\bar{\varphi}^i
\ee
with the shadow state
\be\label{eq36}
\bar{\varphi}^j=\cdots\circ\bar{\phi}_i^j\circ\cdots\bar{\phi}_j^j\circ\cdots\bar{\phi}_k^j\circ\cdots.
\ee
When we detect a particle in zone $i$, i.e. $|\phi_i^i|^2dV_i=1$, the shadow state $ \bar{\varphi}^i$ is determined simultaneously, so is the state $ \varphi^i$. Thus the wave function collapses all over the coordinate space simultaneously through the QV and the shadow states.

The existence  of the QV is beyond  question. It is the QV that causes an atom' spontaneous emission \cite{kvah} and the Casimir effect\cite{mkrg,mbum}. It is well known that there is an effect of the QV polarization by external fields \cite{schw} and that particles can be created from the QV by external fields \cite{miri}. The QV is defined in a Fock space and can be understood easily. But to describe the QV in the coordinate space is very difficult, and the description will look very counterintuitive and inconsistent.    The unintelligibility of the quantum theory is mainly due to the indescribability with the coordinate space of the QV  which, in my point of view, is the deepest mystery in nature. The QV has infinity energy and infinity potential to generate infinity particles. But we can't see it or detect it directly. The QV is the strongest physical reality since nothing can destroy it even the so-called Big Bang, at the same time it is the weakest reality because even the smallest particle can move in it without any obstacle from it. It exists all over the coordinate space, at the same time considering the fact that quantum correlations take place without any time-ordering, it should be considered to be a non-spatial domain where there is no distance  and  as if the space collapses - it is the largest reality in the world because every object is in his body while it is the smallest reality in the world considering that there is no spatial structure in the QV. In this way we can find numerous contradiction in terms to describe the QV.

Any quantum object has the so-called wave particle duality. In the neutron double-slit experiments the intensity of the neutron sources is so low that  while one neutron is being recorded, the next one to be recorded is still confined to the neutron source, leading to the conclusion that the interference fringe is really collected one by one, which shows the particle nature \cite{azei,azrg}.  Where does the wave characteristic of the interference come from? Why does a single particle have wave characteristic?   Since any particle is created from the QV which is nonlocal  and has the characteristic of superposition, at the same time leaving the QV broken with a shadow, which leads to the fact that any particle is inseparable from the QV, it is very reasonable to ascribe the wave nature of any quantum object to the shadowed QV.  Thus the quantum object can be taken as a system  comprising two inseparable parts - a particle and the shadowed  QV, and its state is perfectly described by a wave function which obeys the well-known equations. Bearing this picture in mind, we will not be astounded by the recent experimental advances  \cite{sgtp,cbnb,pate,bran} which show that certain features of realistic descriptions must be abandoned and ``one should not have any realistic pictures in one's mind when considering a quantum phenomenon" \cite{azei}. This is due to the fact that any quantum object's inseparable part - the shadowed QV can't be described by any classical language which we are familiar with.

I have shown that any particle created from the QV leaves the QV with a corresponding shadow state in it. Based on this discovery,  quantum entanglement, quantum teleportation of unknown states, quantum entanglement swap, and wave function collapse in the coordinate can be easily understood.  The interaction between  entangled particles is based on the shadowed QV, is distance-independent, and is instantaneous. I also depict the quantum object as the composite of two inseparate components, a particle and the shadowed QV which the wave nature of the quantum object is ascribed to. The QV is the origin of every particle and even our whole physical world. Assuming the QV is the origin of the consciousness is a natural extend, then the whole cosmos
is created from the QV.  Many scientists such as N. Bohr,  W. Heisenberg, and R. Oppenheimer have realized the close relationship between the modern physics and Eastern mysticism: ``Here the parallels between modern physics and Eastern mysticism are most striking, and we shall often encounter statements where it is almost impossible to say whether they have been made by physicists or by Eastern mystics" \cite{capra}. Now it is clear that this phenomena taken place is due to the fact  that the QV  is also the origin and the core  of Eastern mysticism, in other words, the QV is the body of Buddha and  the Tao which is thought to be the origin of our world. Several hundreds years ago, science said good-bye to religions, today science has arrived at the same destination, the QUANTUM VACUUM, as that  many religions had already reached in a different way.


\section*{Acknowledgements}
This work was supported by the National Nature Science Foundation of China (Grant No. 10874071).



\begin{references}
\bibitem{jfcs}Clauser, J.F. \& Shimony, A., Bell's theorem: Experimental tests and implications. {\it Rep. Prog. Phys.} {\bf 41}, 1881-1927 (1978).
\bibitem{aepb}Einstein, A., Podolsky, B. \& Rosen, N., Can quantum-mechanical description of physical reality be considered complete? {\it Phys. Rev.} {\bf 47}, 777-780 (1935).
\bibitem{jsb} Bell, J.S., On the Einstein-Podolsky-Rosen paradox. {\it Physics} {\bf 1}, 195-200 (1965).
\bibitem{jfch}Clauser, J.F., Horne, M.A.,  Shimony, A. \&  Holt, R.A., Proposed experiment to test local hidden-variable theories. {\it Phys. Rev. Lett.} {\bf 23}, 880-884 (1969).
\bibitem{sjfc}Freedman, S.J. \& Clauser, J.F., Experimental test of local hidden-variable theories.  {\it Phys. Rev. Lett.} {\bf 28}, 938-941 (1972).
\bibitem{esft} Fry, E.S. \& Thompson, R.C., Experimental test of local hidden-variable theories. {\it Phys. Rev. Lett.} {\bf 37}, 465
(1976).
\bibitem{aagp}Aspect, A.,   Grangier, P. \&  Roger, G., Experimental realization of Einstein-Podolsky-Rosen-Bohm gedankenexperiment: a new violation of Bell's inequalities. {\it Phys. Rev. Lett.} {\bf 49}, 91-94
(1982).
\bibitem{zyom} Ou, Z.Y. \&  Mandel, L., Violation of Bell's inequality and classical probability in a two-photon correlation experiment. {\it Phys. Rev. Lett.} {\bf 61}, 50-53 (1988).
\bibitem{prtr}Tapster, P.R.,  Rarity, J.G. \&  Owens, P.C.M., Violation of Bell's inequality over 4 km of optical fiber. {\it Phys. Rev. Lett.} {\bf 73}, 1923-1926
(1994).
\bibitem{pgkm}Kwiat, P.G., Mattle, K.,   Weinfurter, H. \&  Zeilinger, A.,  New high-intensity source of polarization-entangled photon pairs. {\it Phys. Rev. Lett.} {\bf 75}, 4337-4341 (1995).
\bibitem{wtbj}Tittel, W.,  Brendel, J., Zbinden, H \& Gisin, N., Violation of Bell inequalities by photons more than 10 km apart. {\it Phys. Rev. Lett.} {\bf 81}, 3563
(1998).
\bibitem{mard}Rowe, M.A., Kielpinski, D., Meyer, V.,  Sackett, C.A., Itano, W.M.,  Monroe, C. \&  Wineland, D.J., Experimental violation of a Bell's inequality with efficient detection. {\it  Nature} {\bf 409}, 791-794 (2001).
\bibitem{dsab}Salart, D., Baas, A., Branciard, C.,  Gisin, N. \& Zbinden, H., Testing the speed of `spooky action at a distance'. {\it Nature} {\bf 454}, 861-864
(2008).
\bibitem{dmpm} Matsukevich, D.N.,  Maunz, P., Moehring, D.L., Olmschenk, S. \&  Monroe, C., Bell inequality violation with two remote atomic qubits, {\it Phys. Rev. Lett.} {\bf 100}, 150404 (2008).
\bibitem{gisin} Gisin, N., Quantum nonlocality:
how does nature do it? {\it Science} {\bf 326}, 1357-1358 (2009).
\bibitem{dbjp}Bouwmeester, D., Pan, J.-W., Mattle, K., Eibl, M., Weinfurter, H. \& Zeilinger, A.,  Experimental quantum teleportation. {\it Nature} {\bf 390}, 575-579 (1997).
\bibitem{jpdb} Pan, J.-W.,  Bouwmeester, D., Weinfurter, H. \& Zeilinger, A., Experimental entanglement swapping: entangling photons that never interacted. {\it Phys. Rev. Lett.} {\bf 80}, 3891-3894 (1998).
\bibitem{capra} Capra, E.,  {\it The tao of phisics} (Shambhala, Boulder, 1975).
\bibitem{mpds} Peskin, M.E. \& Schroeder, D.V.,  {\it An introduction to quantum field theory} (Westview, Boulder, 1995).

\bibitem{luri} Lurie, D., {\it Particles and fields} (Interscience,  1968).


 \bibitem{ashz} Stefanov, A., Zbinden, H.,   Gisin, N. \& Suarez, A., Quantum correlations with spacelike separated beam splitters in motion: experimental test of multisimultaneity. {\it Phys. Rev. Lett.} {\bf 88}, 120404 (2002).
\bibitem{hzjb} Zbinden, H., Brendel, J.,  Gisin, N.\& Tittel, W., Experimental test of nonlocal quantum correlation in relativistic configurations. {\it Phys. Rev. A} {\bf 63}, 022111 (2001).

 \bibitem{kvah} Vahala, K.J.,   Optical microcavities. {\it Nature} {\bf 424}, 839-846 (2003).
\bibitem{mkrg} Kardar, M., \& Golestanian , R.,  The ``friction"of vacuum, and other fluctuation-induced forces. {\it Rev. Mod. Phys.} {\bf 71}, 1233-1245 (1999).
\bibitem{mbum} Bordag, M.,  Mohideen, U., \& Mostepanenko, V.M.,  New Developments in the Casimir Effect. {\it Phys. Rept.} {\bf 353}, 1-205 (2001).

\bibitem{schw} Schwinger, J.,   On Gauge Invariance and Vacuum Polarization. {\it Phys. Rev.} {\bf 82}, 664-679 (1951).   \bibitem{miri} Riordan, M.,   The discovery of quarks. {\it Science} {\bf 256}, 1287-1293 (1992).





















\bibitem{azei} Zeilinger, A., Experiment and the foundations of quantum physics. {\it Rev. Mod. Phys.} {\bf 71}, S288-S297 (1999).
\bibitem{azrg} Zeilinger, A., G\"{a}hler, R., Shull, C.G., Treimer, W., \& Mampe, W.,  Single- and double-slit diffraction of neutrons. {\it Rev. Mod. Phys.} {\bf 60}, 1067-1073 (1988).







\bibitem{sgtp}Gr\"{o}blacher, S., Paterek, T., Kaltenbaek, R.,  Brukner, \v{C}., \.{Z}ukowski, M.,
 Aspelmeyer, M. \& Zeilinger, A.,  An experimental test of non-local realism. {\it Nature} {\bf 446}, 871-875 (2007).
\bibitem{cbnb} Branciard, C., Brunner, N., Gisin, N., Kurtsiefer, C., Lamas-Linares, A., Ling, A. \& Scarani, V.,   Testing quantum correlations versus single-particle properties within
Leggett's model and beyond. {\it Nat. Phys.} {\bf 4}, 681-685 (2008).
\bibitem{pate}Paterek, T. {\it et al.} Experimental test of nonlocal realistic theories without the rotational symmetry
assumption. {\it Phys. Rev. Lett.} {\bf 99}, 210406 (2007).
\bibitem{bran}Branciard, C. {\it et al.} Experimental falsification of Leggett's nonlocal variable model. {\it Phys. Rev. Lett.} {\bf 99}, 210407 (2007).



\end{references}
\end{document}